\documentclass[aps,pre,reprint,superscriptaddress,showpacs,showkeys,10pt,twocolumn]{revtex4-1}

\usepackage{amsfonts}
\usepackage{amssymb}
\usepackage{amsmath}
\usepackage{graphicx}
\usepackage{epstopdf}
\usepackage[normalem]{ulem}
\usepackage{multirow}
\usepackage{color}

\begin{document} 

\title{Handling magnetic anisotropy and magnetoimpedance effect in flexible multilayers under external stress}

\author{K.~Agra} 
\affiliation{Departamento de F\'{i}sica Te\'{o}rica e Experimental, Universidade Federal do Rio Grande do Norte, 59078-900 Natal, RN, Brazil} 
\author{F.~Bohn} 
\affiliation{Departamento de F\'{i}sica Te\'{o}rica e Experimental, Universidade Federal do Rio Grande do Norte, 59078-900 Natal, RN, Brazil} 
\author{T.~J.~A.~Mori} 
\affiliation{Laborat\'{o}rio Nacional de Luz S\'{i}ncrotron, Rua Giuseppe M{\'a}ximo Scolfaro, 1000, Guar{\'a}, 13083-100 Campinas, SP, Brazil}
\author{G.~L.~Callegari} 
\affiliation{Departamento de F\'{i}sica, Universidade Federal de Santa Maria, 97105-900 Santa Maria, RS, Brazil}
\author{L.~S.~Dorneles} 
\affiliation{Departamento de F\'{i}sica, Universidade Federal de Santa Maria, 97105-900 Santa Maria, RS, Brazil}
\author{M.~A.~Correa} 
\email[Electronic address: ]{marciocorrea@dfte.ufrn.br}
\affiliation{Departamento de F\'{i}sica Te\'{o}rica e Experimental, Universidade Federal do Rio Grande do Norte, 59078-900 Natal, RN, Brazil} 

\date{\today} 

\begin{abstract}
We investigate the dynamic magnetic response though magnetoimpedance effect of ferromagnetic flexible NiFe/Ta and FeCuNbSiB/Ta multilayers under external stress. We explore the possibility of handling magnetic anisotropy, and consequently the magnetoimpedance effect, of flexible magnetostrictive multilayers. We quantify the sensitivity of the multilayers under external stress by calculating the ratio between impedance variations and external stress changes, and show that considerable values can be reached by tuning the magnetic field, frequency, magnetostriction constant, and external stress. The results extend possibilities of application of flexible magnetostrictive multilayers under external stress and place them as very attractive candidates as element sensor for the development of sensitive smart touch sensors.
\end{abstract}



\maketitle 
\section{Introduction}

In the recent years, as consequence of an increasing number of works exploring the integration between magnetic properties and flexible systems~\cite{APL96p072502,APL104p062412,CPB24p077501,JMMM378p499,EPL114p17003}, many applications have been proposed using the fascinating characteristics of ferromagnetic flexible nanostructures. Experiments have been carried out in numerous samples, including ribbons and films. However, although the ribbons may present very interesting properties for flexible applications~\cite{MA182p2411}, they are limited to the electronic integration due to the difficulty in the miniaturization and integralization with lithography techniques. In contrast, thin films, multilayers and flexible substrates overcome these issues and appear as key elements, having striking potential of application in distinct smart devices. For instance, they have been investigated to be applied in electronic skin to mimic the nature with respect to functionality and appearance~\cite{NC6p6080,JAP106p034503,NM12p938, NC4p1859}, as well as have been widely explored as ground for spintronics devices mainly due their magnetic and mechanical properties~\cite{APL107p252401,APL96p072502,APL105p062409,APL104p062412,EPL114p17003}, in a sense that magnetic properties can enable/disable the film applicability in a specific magnetic devices. This explains the recent interest in controlling and handling of properties as magnetic anisotropy, dynamic magnetic response, magnetostrictive properties and stress in ferromagnetic flexible nanostructures.

In this context, the magnetoimpedance effect (MI) corresponds to a powerfull tool to investigate magnetic materials and the dynamic response, as well as it is of technological interest due to the application of materials exhibiting MI in sensor devices~\cite{JMMM242p33}. The MI corresponds to the change of the real and imaginary components of electrical impedance of a ferromagnetic sample caused by the action of an external static magnetic field. In a typical MI experiment, the studied sample is also submitted to an alternate magnetic field associated to the electric current $I_{ac} = I_o \exp (i2 \pi ft)$, $f$ being the probe current frequency. Irrespective to the sample geometry, the overall effect of these magnetic fields is to induce strong modifications of the effective magnetic permeability~\cite{JAP116p243904}. 

Regarding to MI in flexible multilayers, in the recent past, distinct groups have reported very interesting results~\cite{JMMM355p136,JMMM393p593,NMM48p1375,JMMM415p91,NRL7p230,JNN12p7496}, opening the possibilities for the use of flexible substrates in the development of MI based sensors devices for field detection. In particular, our group has shown that non-magnetostrictive NiFe/(Ag,Ta) multilayers with similar magnetic properties and dynamic magnetic response can be obtained in distinct substrates~\cite{JMMM355p136}. For magnetostrictive multilayers, we have performed similar study in Co/(Ag,Cu,Ta) structure~\cite{JMMM393p593}, and the mirroring of the magnetic features considering rigid and flexible substrates has been also verified~\cite{JMMM393p593}. This corresponds to a fundamental issue, since the optimized response obtained for films in ordinary rigid substrates can be reproducible in flexible ones~\cite{JMMM355p136,JMMM393p593}. However, considering both aforementioned reports, it is important to emphasize that, up to now, we have investigated the magnetic properties just in multilayers without external stress. 

In this work, we report on the magnetoimpedance effect in ferromagnetic flexible NiFe/Ta and FeCuNbSiB/Ta multilayers under external stress. We handle magnetic anisotropy and magnetoimpedance effect of the flexible magnetostrictive nanostructures and verify that the sensitivity of the multilayers under external stress, calculated by the ratio between impedance changes and external stress variations, reaches considerable values and is experimentally tunable by the magnetic field, frequency, magnetostriction constant, and external stress. The results extend possibilities of application of flexible magnetostrictive multilayers under external stress and place them as very attractive candidates as element sensor for the development of sensitive smart touch sensors.

\section{Experimental procedure}

For this study, we produce ferromagnetic flexible multilayers with distinct magnetostrictive properties. We select ferromagnetic alloys (FM) with nominal composition of Ni$_{81}$Fe$_{19}$, the well-known Permalloy which has vanishing magnetostriction~\cite{PR28p146}, and Fe$_{73.5}$Cu$_{1}$Nb$_{3}$Si$_{13.5}$B$_{9}$, the precursor of the so-called nanocrystalline Finemet~\cite{JMMM288p347} that presents high positive saturation magnetostriction constant~\cite{JMMM112p258}, $\lambda_{s} \approx + 26 \times 10^{-6}$, and soft magnetic properties even in the amorphous state~\cite{JPDAP41p175003}. The [FM($10$ nm)/Ta($2$ nm)]$\times 50$ multilayers are deposited by magnetron sputtering onto flexible Kapton\textsuperscript{\textregistered} substrate, covered with a $2$~nm-thick Ta buffer layer, with dimensions of around $10\times 4$~mm$^2$, using the following parameters: base pressure of $7 \times 10^{-7}$~Torr, deposition pressure of $2 \times 10^{-3}$ Torr with Ar at $32$~sccm constant flow, and using a DC source with $20$~W. Under these conditions, the deposition rates for the NiFe, FeCuNbSiB and Ta layers are $0.52$~nm/s, $0.33$~nm/s and $0.18$~nm/s, respectively. During the film deposition, the substrate moves at constant speed through the plasma to improve the film uniformity, and a constant $1$ kOe magnetic field is applied perpendicularly to the main axis of the substrate to induce magnetic anisotropy and define an easy magnetization axis. For comparision, multilayers deposited onto rigid glass substrate are also produced in the same batch. In particular, these multilayer are used just to make a comparision with respect to the structural characterization (not shown). 

The structural characterization of the multilayers is performed through x-ray reflectometry (XRR) and x-ray diffraction (XRD) performed with a Rigaku-Miniflex diffractometer in a Bragg-Brentano geometry and using CuK$_\alpha$ radiation.

The magnetic properties and magnetization dynamics are investigated in ferromagnetic flexible multilayers with distinct magnetostrictive properties under external stress. Experimentally, the multilayers are submitted to external stress by bending the substrate with predefined curvatures.  

\begin{figure}[!h]
\vspace{-0.2cm}
\begin{center}
\includegraphics[width=8.5cm]{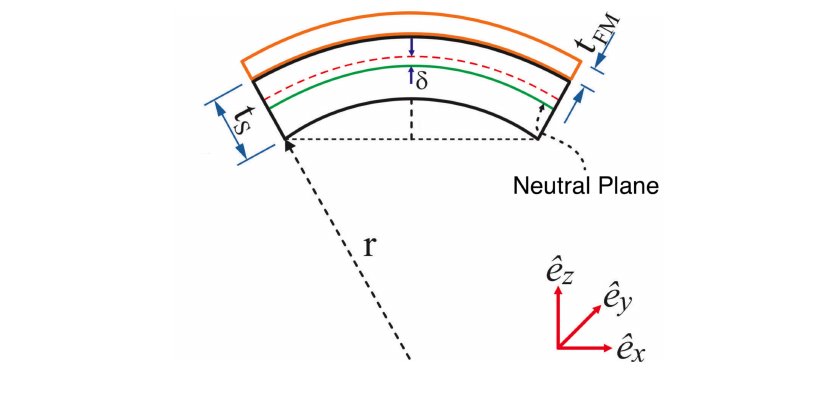}
\end{center}
\vspace{-.7cm}
\caption{Experimental parameters considered for the estimation of the external stress. In this case, $\delta$ is the displacement of the neutral plane, $r$ the curvature radius in which the multilayer is submitted, and $t_{FM}$ and $t_{S}$ are the thicknesses of the film and of the substrate, respectively.}
\label{fig01}
\end{figure}

The external stress in the multilayer may be estimated following the approach described in Refs.~\cite{JAP91p9652,JAP96p4154}, which is based on the Stoney Model and considers the bending of the sample. To this end, some assumptions are taken into account: ({\it i}) The film must be planar, homogeneous and isotropic; ({\it ii}) The thicknesses of the film ($t_{FM}$) and of the substrate ($t_{S}$) must be uniform; and ({\it iii}) $t_{FM} \ll t_{S}$. For the multilayers studied here, $t_{FM} \approx 0.0046\, t_{S}$. From the Hooke's Law, where the Young's modulus $Y$ and Poisson ratio $\nu_{s}$ are considered, the average stress $\bar{\sigma}$ along the main axis of the sample can be written as
\begin{equation}
\bar{\sigma}=\frac{Y}{6(1-\nu_{s})r}\frac{t_{S}^2}{t_{FM}},
\end{equation}
\noindent where $r$ is the curvature radius, that has a dependence with the sample's dimensions. Figure~\ref{fig01} presents the experimental parameters considered for the estimation of the external stress. Thus, considering that our samples present small differences of size, the estimated average stress is distinct for each induced bending. At the same time, the multilayers present a natural curvature, leading to a compressive stress when the curvature is reduced, a fact primarily evidenced for the FeCuNbSiB/Ta multilayers. Here, we consider the selected $\bar{\sigma}$ values of $32.1$, $64.7$, $92.0$, and $124.2$~MPa for the NiFe/Ta multilayer and $-1.5$, $44.1$, $82.4$, and $127.3$~MPa for the FeCuNbSiB/Ta one. 

The in-plane magnetic properties are obtained through magnetization curves measured along and perpendicularly the main axis of the samples with a Lakeshore $7400$ Vibrating Sample Magnetometer. All the curves are obtained with maximum applied magnetic field of $\pm 350$ Oe at room temperature.

The magnetoimpedance effect is measured using a RF-impedance analyzer Agilent model $E4991$, with $E4991A$ test head connected to a microstrip in which the sample is the central conductor~\footnote{See supplementary material for a movie showing an illustration of the MI experiment for a flexible magnetostrictive multilayer under external stress.}, which is separated from the ground plane by the substrate. The electric contacts between the sample and the sample holder are made with $24$ h cured low resistance silver paint. To avoid propagative effects and acquire just the sample contribution to MI, the RF impedance analyzer is calibrated at the end of the connection cable by performing open, short, and load ($50$ $\Omega$) measurements using reference standards. The probe current is fed directly to one side of the sample, while the other side is in short circuit with the ground plane. The {\it ac} current and external magnetic field are applied along the length of the sample. MI measurement is taken over a wide frequency range, between $0.5$ GHz and $3.0$ GHz, with maximum applied magnetic fields of $\pm 350$ Oe. While the external magnetic field is swept, a $0$ dBm ($1$ mW) constant power is applied to the sample characterizing a linear regime of driving signal. Thus, at a given field value, the frequency sweep is made and the real $R$ and imaginary $X$ parts of the impedance $Z$ are simultaneously acquired. 

\section{Results}

\begin{figure}[!h]
\vspace{-0.2cm}
\begin{center}
\includegraphics[width=8.5cm]{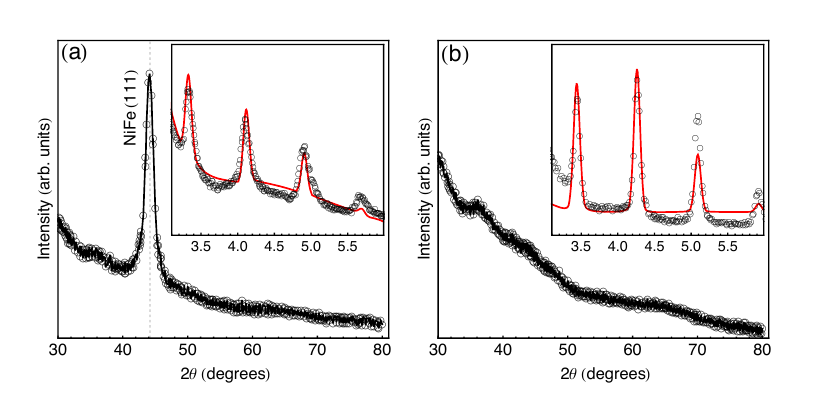}
\end{center}
\vspace{-.7cm}
\caption{X-ray diffraction and X-ray reflectometry (insets) for the (a) NiFe/Ta and (b) FeCuNbSiB/Ta multilayers. In the insets, the symbols correspond to the experimental data, while the solid lines represent the simulations for the XRR.}
\label{fig02}
\end{figure}

Figure~\ref{fig02} shows the XRD and XRR patterns obtained for both the NiFe/Ta and FeCuNbSiB/Ta multilayers grown on flexible substrates. The presence of only one peak for the NiFe/Ta multilayer, presented in the XRD diffractogram, indicates that the NiFe layers are well-textured with the (111) planes parallel to the surface. The NiFe crystallite size estimated by the Debye-Scherrer equation is within the error of the layer thickness. On the other hand, FeCuNbSiB does not present long range crystalline order as its diffractogram shows just a very broad peak around 2$\theta \approx$ $44^\circ$. The XRR patterns confirm the flatness and homogeneity of the two samples. By simulating the data was found average roughnesses as small as $0.66$ nm ($0.50$ nm) for the Tantalum surfaces in the NiFe/Ta (FeCuNbSiB/Ta) multilayer, and $0.69$ nm ($0.67$ nm) for the NiFe (FeCuNbSiB) surfaces. The exact thicknesses are 2.54 nm for Ta layers (both multilayers), $8.41$ nm and $7.98$ nm for NiFe and  FeCuNbSiB, respectively. Reference samples grown at the same batch, but on rigid glass substrates instead, were found to be about $1.7$ $\%$ thicker with respect to the ones discussed here. This strain may be related to the releasing of residual stress which is made possible by the flexibility of the Kapton\textsuperscript{\textregistered} substrate and leads to the natural curvature mentioned previous section.

\begin{figure}[!h]
\begin{center}
\includegraphics[width=8.5cm]{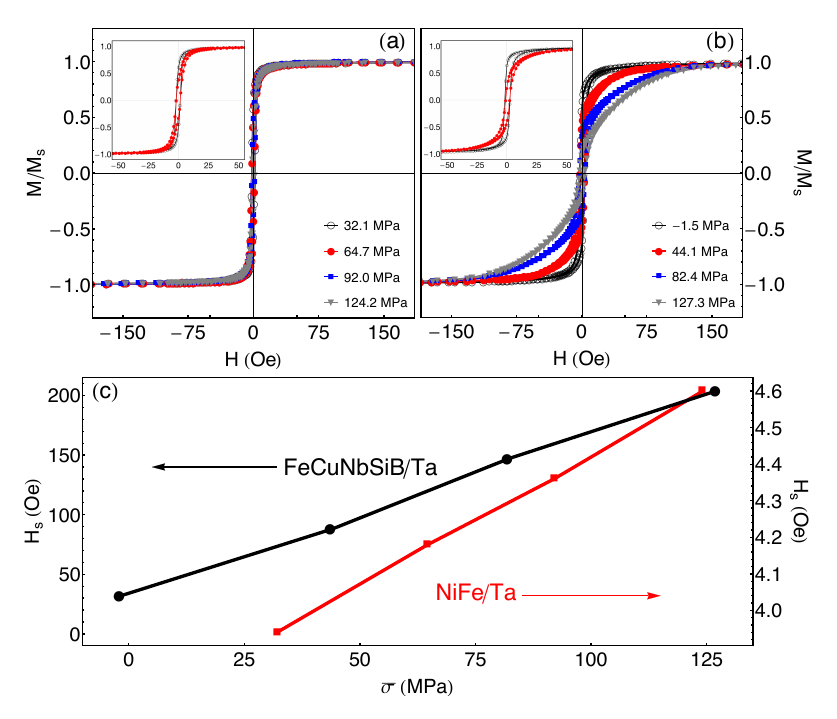}
\end{center}
\vspace{-.8cm}
\caption{Normalized magnetization curves for the (a) NiFe/Ta multilayer and (b) FeCuNbSiB/Ta one under selected values of external stress. Notice that the curves are obtained with the magnetic field perpendicular to the main axis of the samples, while the external stress is applied along the main axis. The insets present normalized magnetization curves measured along (filled symbols) and perpendicularly (open symbols) the main axis for the multilayers without external stress. (c) Saturation field as a function of external stress for both multilayers.}
\label{fig03}
\end{figure}

Figure~\ref{fig03}(a,b) shows the normalized magnetization curves obtained for the multilayers under the different external stress values, while Fig.~\ref{fig03}(c) presents the stress dependence of the saturation field $H_s$, obtained from the magnetization curves. Soft magnetic properties are verified for both multilayers. The curves obtained for the multilayers without external stress, {\it i.\ e.}, under residual stress, confirm an in-plane magnetic anisotropy, with easy magnetization axis perpendicular to the main axis, induced by the magnetic field applied during the deposition, as expected. However, as the external stress is increased, it is identified a clear difference of the magnetic behavior of the multilayers, a fingerprint of the distinct magnetostriction values. 

It is important to point out that the external stress can give rise significant changes in the magnetic properties, magnetic domain structure and magnetization processes of the multilayers. The magnetoelastic anisotropy energy term associated to the external stress can be expressed by $E_\sigma = \frac{3}{2}\lambda_S\sigma$. In particular, the magnetoelastic anisotropy induced by external stress competes by enhancing or balancing anisotropies induced during the production process of the multilayers. Notice that the result of the applied external stress is strongly dependent on the product between $\lambda$ and $\sigma$. Thus, if $\lambda\sigma>0$, an anisotropy axis of magnetoelastic origin is induced along the same direction of the applied external stress, otherwise, if $\lambda\sigma<0$, this anisotropy axis is oriented perpendicularly to the direction of application of the external stress~\cite{Cullity}.

The curves for the NiFe/Ta multilayer do not present any dependence with the applied stress and a very tiny variation of $H_s$ is verified, signatures of the vanishing magnetostriction of the investigated NiFe alloy. On the other hand, for the FeCuNbSiB/Ta multilayer, the magnetic properties are strongly dependent on the external stress. It is observed an evolution of the shape of the magnetization curve, from a squared shaped loop to an inclined and narrowed curve, as well as a large variation of $H_s$ with the external stress. Since the FeCuNbSiB alloy has high positive magnetostriction, a magnetoelastic anisotropy axis arises along the direction of the positive external stress, {\it i.\ e.}, along the main axis of the multilayer, which leads to a change in the orientation of the effective magnetic anisotropy with the increase of the stress value. In this case, the magnetic behavior is a result of the competition between the induced uniaxial magnetic anisotropy and the magnetoelastic anisotropy contribution. In this sense, this competition allows us to handle the magnetic anisotropy and the magnetoimpedance effect of ferromagnetic flexible substrates.

The quasi-static magnetic properties play a fundamental role in the dynamic magnetic response of a ferromagnetic system. Thus, the changes in the magnetic anisotropy will be reflected in the magnetoimpedance curves. Regarding the results, in order to allow direct comparison between the results, we define $\textrm{MI} = Z (H) - Z(H_{max})$, where $Z(H)$ is the impedance at a given external magnetic field and $Z(H_{max})$ is the impedance for the maximum external magnetic field applied, where the multilayer is saturated magnetically. 

\begin{figure}[!h]
\begin{center}
\includegraphics[width=8.5cm]{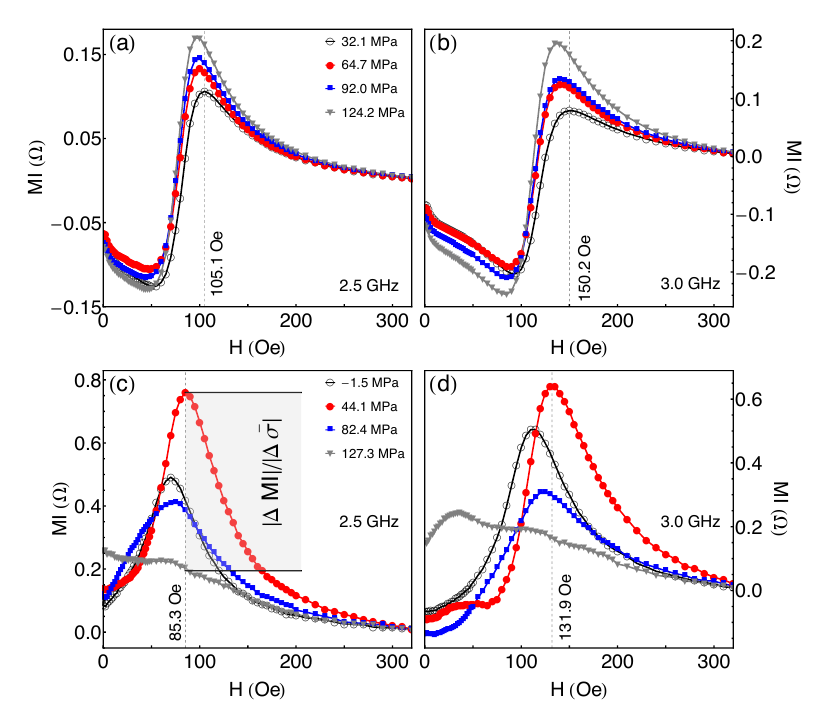}
\end{center}
\vspace{-.8cm}
\caption{The MI curves as a function of the external magnetic field for the NiFe/Ta multilayer at the frequencies of (a) $2.5$ and (b) $3.0$ GHz under different external stress values. (c,d) Similar plot for the FeCuNbSiB/Ta multilayer. Although the curves are acquired over a complete magnetization loop and present hysteretic behavior, we show just the part of the curve with increasing field and positive values to make clearer the visualization. The dashed lines indicate the location in field of the MI peak, named $H_{ref}$, obtained at the respective frequency and when each multilayer is under the smaller positive $\bar{\sigma}$ value, $32.1$ and $44.1$ MPa, respectively. In (c), the gray zone indicates the MI values employed for the estimation of the sensitivity of the multilayer under stress.}
\label{fig04}
\end{figure}

Figure~\ref{fig04} shows the MI curves, at selected frequencies, for the multilayers under different external stress values. It is known that the shape and amplitude of the MI curves are strongly dependent on the orientation of the magnetic field and {\it ac} current with respect to the magnetic anisotropies~\cite{APL67p857, APL104p102405}, magnitude of the magnetic field, probe current frequency, as well as are directly related to the main mechanisms responsible for the transverse magnetic permeability changes: skin effect and ferromagnetic resonance (FMR) effect~\cite{JAP110p093914}. Here, the well-known symmetric magnetoimpedance behavior around $H = 0$ for anisotropic systems~\cite{APL67p857} is verified and the MI curves reflect all the expected classical features, including the MI behavior with respect to the orientation between the anisotropy and field~\cite{SAA106p187}, as well as with frequency~\cite{SAA106p187,JMMM195p764,APL69p3084,JAP110p093914}. 

The NiFe/Ta multilayer exhibits curves with a double peak behavior for the whole frequency range, a signature of the perpendicular alignment of the field and {\it ac} current with the easy magnetization axis. An interesting feature resides in the dependence of the MI amplitude and peaks position with the probe current frequency and the external stress. For frequencies up to $\sim 0.7$ GHz, not shown here, the peaks position remains close to the anisotropy field, indicating the skin effect as the main responsible for the changes of the transverse permeability. For frequencies above this value, the MI variation is a consequence of the a strong skin and FMR effects acting simultaneously in the multilayers, a fact evidenced by the displacement of the peaks position towards higher fields as the frequency increases. 
Moreover, the MI results corroborate the negligible influence of the stress on the magnetic properties and on dynamic magnetic response due the vanishing magnetostriction.

On the other hand, for the FeCuNbSiB/Ta multilayer, the peaks structure is strongly affected by the external stress. The evolution in the shape, amplitude and position of the MI peaks is a result of the modifications of the magnetic properties, {\it i.\ e.}, the change in the intensity of the effective magnetic anisotropy occurring with the increase of the external stress value, and its orientation with respect to the direction of the magnetic field and {\it ac} current. The changes of the effective anisotropy lead to variations of the frequency ranges in which the distinct mechanisms governing the dynamic magnetic response act~\cite{JAP116p243904}. The well-defined double peak behavior is verified in the whole frequency range for negative external stress, as well as for small positive stress values, in which the effective anisotropy is perpendicular to the main axis. For these situations, the FMR starts appearing above $\sim 1.7$ GHz. As the stress increases and the orientation of the effective magnetic anisotropy changes, this double peak behavior gives place to a single peak structure for frequencies below $\sim 2.3$ GHz, where the skin effect is the main responsible for the changes of the transverse permeability. For frequencies above this value, the double peak structure is retrieved, a consequence of the a strong skin and FMR effects acting simultaneously in the multilayers.

By considering magnetostrictive multilayers, the best MI response can be explored by playing with sample's parameters, such as substrate, type of structure, thickness and composition of the ferromagnetic material, the latter directly related to magnetic properties as magnetostriction, and experimental parameters. Thus, the most striking finding here resides in the MI variations tuned by modifying the magnetic field, frequency, magnetostriction, and external stress.

\begin{figure}[!b]
\vspace{-.4cm}
\begin{center}
\includegraphics[width=8.5cm]{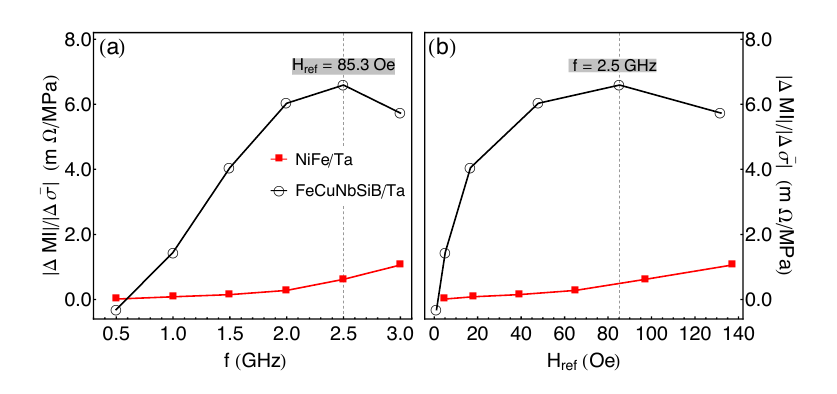}
\end{center}
\vspace{-.8cm}
\caption{The $|\Delta \textrm{MI}|/|\Delta \bar{\sigma}|$ ratio as a function of (a) frequency~$f$ and (b) magnetic field~$H_{ref}$, indicating the sensitivity for each multilayer under external stress. The dashed lines relate, in distinct plots, the ratio calculated for the FeCuNbSiB/Ta multilayer at frequency of $2.5$~GHz, in which $H_{ref}=85.3$~Oe, corresponding to the highest obtained sensitivity.}
\label{fig05}
\end{figure}

To quantify the sensitivity of the multilayers under external stress, we calculate the magnitude of the MI variations at a given frequency with the change of the stress through
\begin{equation}
\frac{\Delta \textrm{MI}}{\Delta \bar{\sigma}} = \frac{|\textrm{MI}-\textrm{MI}_{ref}|}{|\bar{\sigma}-\bar{\sigma}_{ref}|}.
\label{eq_sens}
\end{equation}
For convenience, as reference quantities, $\textrm{MI}_{ref}$ and $\bar{\sigma}_{ref}$, we consider the MI peak value, which is located in a field $H_{ref}$ identified in the experiment, when the multilayer is under the {\it smaller positive} $\bar{\sigma}$ value. Table~\ref{tab01} presents the $H_{ref}$ values at selected frequency values. The following quantities in the equation correspond simply to the MI value, {\it at the same} $H_{ref}$ and $f$ ones, obtained when the multilayer is under a given $\bar{\sigma}$. Here, we consider the respective data obtained for the {\it higher positive} $\bar{\sigma}$ values for each multilayer, as previously indicated in Fig.~\ref{fig03}.
\begin{table}[!hb]
\vspace{-0.2cm}
\centering 
\caption{Location in field of the MI peak at selected frequency values for the NiFe/Ta and FeCuNbSiB/Ta multilayers, obtained from the curves acquired with each multilayer under the smaller positive $\bar{\sigma}$ values, $32.1$ and $44.1$ MPa, respectively.}
\label{tab01}
\begin{tabular}{ccccc} \hline\hline
$f$ (GHz)         &&  \multicolumn{3}{c}{{ \centering $H_{ref}$ (Oe)}}    \\  
\cline{1-1} \cline{3-5}
 &\hspace{.05cm}{\color{white}.}&  NiFe/Ta &\hspace{.3cm}{\color{white}.}&  FeCuNbSiB/Ta \\
\cline{1-1} \cline{3-5}
$0.5$               &&\hspace{0.2cm}  $7.9$    &&  $1.1$ \\
$1.0$               && \hspace{0.2cm}  $20.7$  && $5.1$ \\ 
$1.5$               && \hspace{0.2cm}  $42.4$  &&  $16.7$\\ 
$2.0$               && \hspace{0.2cm}  $70.2$  &&  $47.9$\\ 
$2.5$               && \hspace{0.2cm}  $105.1$  &&  $85.3$\\ 
$3.0$               && \hspace{0.2cm}  $150.2$  &&  $131.9$\\ \hline \hline
\end{tabular}
\label{table1}
\vspace{-0.0cm}
\end{table}
Figure~\ref{fig05} shows the $|\Delta \textrm{MI}|/|\Delta\bar{\sigma}|$ ratio, as defined by Eq.~(\ref{eq_sens}), as a function of frequency $f$ and magnetic field $H_{ref}$, indicating the sensitivity for each multilayer under external stress. From the figure, we clearly verify the strong influence of the magnetostriction on the sensitivity. As expected, the NiFe/Ta multilayer present very low sensitivity due to the reduced differences of the MI values for distinct stress values. A discrete increase is noticed above $2.0$ GHz, since the FMR effect is able to split the MI curves obtained in different conditions, even the magnetic anisotropy is similar in both cases. For the FeCuNbSiB/Ta multilayer, it can be seen that the sensitivity values become significant from $0.6$ GHz. The highest sensitivity is observed at $2.5$ GHz, with $H_{ref}$ of $85.3$ Oe, reaching $\sim 6.6$ m$\Omega$/MPa.

Our results raise an interesting issue on the MI behavior and its application in technological devices based on the effect in flexible magnetostrictive multilayers. Obviously, they can be widely employed in magnetic field sensor devices, in which the field detection is performed through a change in the impedance of the sensor element. However, the flexible magnetostrictive multilayers appear as very attractive candidates for application as probe element to the development of sensitive smart touch sensors. To illustrate it, we can figure out a simple device based on the MI effect, in which the flexible magnetostrictive multilayers works as a logic key. In this case, once the dynamic magnetic response is known and device parameters as $f$ and $H_{res}$ are set, the change of MI, {\it i.~e.}, {\it high} and {\it low} impedance values, is obtained through the bending or external stress applied to the multilayer~\footnote{See supplementary material for a movie showing an illustration of a logic operation device based on MI changes of a flexible magnetostrictive multilayer under external stress, such as a sensitive smart touch sensor.}.

\section{Conclusion}

In conclusion, we have investigated the MI effect in ferromagnetic flexible NiFe/Ta and FeCuNbSiB/Ta multilayers under external stress. We have handled magnetic anisotropy, and consequently the magnetoimpedance effect, of these flexible multilayers. From the results, we have observed that the magnetic properties of the NiFe/Ta multilayer are not considerably influenced by the external stress, as expected due to the vanishing magnetostriction of the ferromagnetic alloy. At the same time, we have verified a strong evolution of the MI response as a results of the modifications of the effective magnetic anisotropy occurring with the increase of the external stress. Thus, we have tuned the sensitivity of the multilayers under external stress by setting the magnetic field, frequency, magnetostriction constant, and external stress. For the studied multilayers, the highest sensitivity is observed for the FeCuNbSiB/Ta one, reaching $6.6$~m$\Omega$/MPa at $2.5$ GHz and $85.3$ Oe. In this sense, the results infer that the flexible magnetostrictive multilayers under external stress are not just suitable for the development of sensors devices for field detection, but also extend possibilities of application, placing them as very attractive candidates as element sensor for the development of sensitive smart touch sensors.

\begin{acknowledgments}
The research is supported by the Brazilian agencies CNPq (Grant nos. 306362/2014-7, 441760/2014-7, 306423/2014-6 and 471302/2013-9), CAPES, and FAPERN (Pronem No. 03/2012). M.A.C. and F.B. acknowledge financial support of the INCT of Space Studies.
\end{acknowledgments}

\end{document}